\documentclass[11pt]{article}
\usepackage{amsmath,amsxtra}
\usepackage{amscd}
\usepackage[dvips]{graphicx}
\usepackage{graphicx}
\usepackage{latexsym}

\usepackage{latexsym}
\usepackage{amsmath}
\date{\nonumber}
\begin{document}
\date{}
\newtheorem{theorem}{Theorem}
\newtheorem{proposition}{Proposition}
\newtheorem{lemma}{Lemma}
\newtheorem{definition}{Definition}

\renewcommand{\theequation}
{\arabic{section}.\arabic{equation}}

\renewcommand{\theequation}
{\arabic{section}.\arabic{equation}}
\title{On Camassa-Holm equation with self-consistent sources and its
solutions}
\author{ Yehui Huang\footnote{Corresponding author: Yehui Huang, Tel:
+86-13810446869, e-mail: huangyh@mails.tsinghua.edu.cn}, Yuqin
Yao\footnote{yuqinyao@mail.tsinghua.edu.cn} and Yunbo
Zeng\footnote{yzeng@math.tsinghua.edu.cn}
\\ Department of Mathematical Sciences,\\ Tsinghua University, Beijing, 100084, P.R. China\\
} \maketitle
\begin{abstract}
Regarded as the integrable generalization of Camassa-Holm (CH)
equation, the CH equation with self-consistent sources (CHESCS) is
derived. The Lax representation of the CHESCS is presented. The
conservation laws for CHESCS are constructed. The peakon solution,
N-soliton, N-cuspon, N-positon and N-negaton solutions of CHESCS are
obtained by using Darboux transformation and the method of variation
of constants.
\end{abstract}

 \vskip .1cm

\section{Introduction}
\setcounter{equation}{0}Camassa-Holm (CH) equation, which was
implicitly contained in the class of multi-Hamiltonian system
introduced by Fuchssteiner and Fokas $^{1}$ and explicitly derived
as a shallow water wave equation by Camassa and Holm $^{2,3}$, has
the form
\begin{equation}
u_t+2\omega u_x-u_{xxt}+3uu_x=2u_xu_{xx}+uu_{xxx},
\end{equation}
where $u=u(x,t)$ is the fluid velocity in the $x$ direction and the
constant $2\omega$ is related to the critical shallow water wave
speed. Let $q=u-u_{xx}+w$, we have the following equivalent equation
$^{4}$.
\begin{equation}\label{eq:CH}
q_t+2u_xq+uq_x=0.
\end{equation}
It was shown by Camassa and Holm that this equation shares most of
the properties of the integrable system of KdV type $^{2,3}$. It
possesses Lax pair formalism and the bi-hamiltonian structure. When
$w>0$, the CH equation has smooth solitary wave solutions. When
$w\longrightarrow0$, these solutions become piecewise smooth and
have cusps at their peaks. These kind of solutions are weak
solutions of (\ref{eq:CH}) with $\omega=0$ and are called "peakons".
Since the works of Camassa and Holm, this equation has become a
well-known example of integrable systems and has been studied from
many kinds of views $^{4-12}$.

Soliton equations with self-consistent sources (SESCS) have
attracted much attention in recent years. They are important
integrable models in many fields of physics, such as hydrodynamics,
state physics, plasma physics, etc $^{13-25}$. For example, the KdV
equation with self-consistent sources describes the interaction of
long and short capillary-gravity waves $^{13}$. The nonlinear
Schr$\ddot{o}$dinger equation with self-consistent sources
represents the nonlinear interaction of an electrostatic
high-frequency wave with the ion acoustic wave in a two component
homogeneous plasma $^{18}$. The KP equation with self-consistent
sources describes the interaction of a long wave with a short wave
packet propagating on the $x$-$y$ plane at some angle to each other
$^{15}$. The SESCS were firstly studied by Melnikov $^{13-15}$. A
systematic way to construct the soliton equations with
self-consistent sources and their zero-curvature representations is
proposed $^{21-24}$.  The problem of finding soliton solutions or
other specific solutions for SESCS has been considered in the past
by many authors $^{13-25}$.

The present paper falls in that line of the work on the CH equation
concerning with establishing the many facts of its completely
integrable character, aiming at the integrable generalization of CH
equation by deriving the Camassa-Holm equation with self-consistent
sources (CHESCS) and finding its solutions. We first construct the
CHESCS by using the approach presented in the reference $^{21-24}$.
The Lax pair of the CHESCS is obtained, which means that the CHESCS
is Lax integrable and can be viewed as integrable generalization of
CH equation. Since the CH equation describes shallow water wave and
the SESCSs in general describe the interaction of different solitary
waves, it is reasonable to speculate on the potential application of
CHESCS, that is, CHESCS may describe the interaction of different
solitary waves in shallow water. It was pointed out $^{26,27}$ that
SESCS can be regarded as soliton equations with non-homogeneous
terms, and accordingly proposed to look for explicit solutions by
using the method of variation of constants. Applying this technique
to CHESCS we have been able to find its peakon solution. In order to
find other solutions of CHESCS, we consider the reciprocal
transformation $^{28,29}$, which relates CH equation to an
alternative of the associated Camassa-Holm (ACH) equation, and
propose the reciprocal transformation, which relates the CHESCS to
associated CHESCS (ACHESCS). By using the Darboux transformation
(DT), one can find the n-soliton and n-cuspon solution $^{8,9}$ as
well as positon and negaton solution of alternative ACH equation.
Then by means of the method of variation of constants, we can obtain
the N-soliton, N-cuspon, N-positon and N-negaton solution for
ACHESCS. Finally, using the inverse reciprocal transformation, we
obtain the N-soliton, N-cuspon, N-positon and N-negaton solution of
CHESCS.

This paper is organized as follows. In section 2, we present how to
derive the CHESCS and its Lax representation. In section 3, the
conservation laws of the CHESCS are constructed. In section 4, the
peakon solution is obtained. In section 5, we consider the
reciprocal transformation for CH equation and CHESCS, respectively.
In section 6, by using the DT, we find the solution for alternative
ACH equation, then by using the method of variation of constants and
inverse reciprocal transformation, we obtain the N-soliton,
N-cuspon, N-positon and N-negaton solution for CHESCS. In section 7,
the conclusion is presented.

\section{The CHESCS and its Lax pair}
\subsection{The CHESCS}
\setcounter{equation}{0} The Lax pair for CH equation (\ref{eq:CH})
is given by $^{2}$
\begin{subequations}
\begin{eqnarray}
\varphi_{xx}&=&(\lambda q+\frac{1}{4})\varphi,\\
\varphi_t&=&(\frac{1}{2\lambda}-u)\varphi_x+\frac{1}{2}u_x\varphi.
\end{eqnarray}
\end{subequations}
It is not difficult to find that
\begin{equation}
\frac{\delta\lambda}{\delta q}=-\lambda\varphi^2.
\end{equation}
The CH equation possesses bi-hamiltonian structure $^{2}$
\begin{equation}\label{eq:CHH}
q_t=-J\frac{\delta H_0}{\delta q}=-K\frac{\delta H_1}{\delta q},
\end{equation}
where
\begin{eqnarray}
K&=&-\partial^3+\partial,\notag\\
J&=&\partial q+q\partial,\notag\\
H_0&=&\frac{1}{2}\int{u^2+u_x^2dx},\notag\\
H_1&=&\frac{1}{2}\int{u^3+uu_x^2dx}.\notag
\end{eqnarray}

According to the approach proposed in the reference $^{21-24}$, the
CHESCS is defined as follows
\begin{subequations}
\begin{eqnarray}
q_t&=&-J(\frac{\delta H_0}{\delta
q}-2\sum_{j=1}^N\frac{\delta\lambda_j}{\delta q})\notag\\
&=&-(q\partial+\partial
q)(u+2\sum_{j=1}^N\lambda_j\varphi_j^2)\notag\\
&=&-2qu_x-uq_x+\sum_{j=1}^N(-8\lambda_jq\varphi_j\varphi_{jx}-2
\lambda_jq_x\varphi_j^2),\\
\varphi_{j,xx}&=&(\lambda_j q+\frac{1}{4})\varphi_j,\quad
j=1,\cdots,N,
\end{eqnarray}
\end{subequations}
which has a equivalent form by using (2.4b)
\begin{subequations}
\begin{eqnarray}
q_t&=&-2qu_x-uq_x+\sum_{j=1}^N[(\varphi_j^2)_x-(\varphi_j^2)_{xxx}],\\
\varphi_{j,xx}&=&(\lambda_j q+\frac{1}{4})\varphi_j,\quad
j=1,\cdots,N,
\end{eqnarray}
\end{subequations}

\subsection{The Lax representation of the CHESCS}
Based on the Lax pair of the CH equation (2.1), we may assume the
Lax representation of the CHESCS (2.4) or (2.5) has the form
\begin{subequations}
\label{eqns:lax4}
 \begin{align}
&\varphi_{xx}=(\lambda q+\frac{1}{4})\varphi,\\
& \varphi_{t}=-\frac{1}{2}B_{x}\varphi+B\varphi_{x},\\
&B=\frac{1}{2\lambda}-u+\sum\limits_{j=1}^{N}\frac{\alpha_{j}f(\varphi_{j})}{\lambda-\lambda_{j}}+
\sum\limits_{j=1}^{N}\beta_{j}f(\varphi_{j}),
 \end{align}
\end{subequations}
where  $f(\varphi_{j})$ is undetermined function of $\varphi_{j}$.
The compatibility condition of (2.6a) and (2.6b) gives
\begin{equation}
\label{eqns:lax5}
\lambda q_{t}=LB+\lambda(2B_{x}q+Bq_{x}),
\end{equation}
where $L=-\frac{1}{2}\partial^{3}+\frac{1}{2}\partial$. Then (2.6)
and (\ref{eqns:lax5}) yields
$$\lambda q_{t}=-\frac{1}{2}\sum\limits_{j=1}^{N}\frac{\alpha_{j}}
{\lambda-\lambda_{j}}[f^{'''}\varphi_{jx}^{3}+3(f^{''}\varphi_{j}-f')(\lambda_{j}q+
\frac{1}{4})\varphi_{jx}+\lambda_{j}q_{x}(f'\varphi_{j}-2f)]$$
$$+[-2qu_{x}-uq_{x}+\sum
\limits_{j=1}^{N}\beta_{j}(2q\varphi_{jx}f'+q_{x}f)]\lambda-\frac{1}{2}
\sum\limits_{j=1}^{N}\beta_{j}[f^{'''}\varphi_{jx}^{2}+(3f^{''}\varphi_{j}+f')$$
\begin{equation}
\label{eqns:expand} \times (\lambda_{j}q+
\frac{1}{4})\varphi_{jx}+\lambda_{j}f'q_{x}\varphi_{j}-f'\varphi_{j}]+\sum
\limits_{j=1}^{N}\alpha_{j}(q_{x}f+2qf'\varphi_{jx}).
\end{equation}
Here $f'$ denotes the partial derivative of the function $f$ with
respect to the variable $\varphi_{j}$. In order to determine
$f,~\alpha_{j}$ and $\beta_{j}$, we compare the coefficients of
$\frac{1} {\lambda-\lambda_{j}},~\lambda$ and $\lambda^{0}$,
respectively. We first observe the coefficients of $\frac{1}
{\lambda-\lambda_{j}}$, then the coefficients of $\varphi_{jx}^{3}$,
$\varphi_{jx}$ and other terms gives rise to, respectively
$$f^{'''}=0,\quad f^{''}\varphi_{j}-f'=0,\quad f'\varphi_{j}-2f=0,$$
which leads to $f=b\varphi_{j}^{2}$. Substituting
$f=b\varphi_{j}^{2}$ into the coefficients of $\lambda$ in (2.8)
gives
$$q_{t}=-2qu_{x}-uq_{x}+4q\sum
\limits_{j=1}^{N}\beta_{j}b\varphi_{j}\varphi_{jx}+q_{x}\sum
\limits_{j=1}^{N}\beta_{j}b\varphi_{j}^{2}.$$ Comparing the above
equation and (2.4a), we can determine
$$b=-2,~\beta_{j}=\lambda_{j}.$$
Substituting $f=-2\varphi_{j}^{2},$ and $\beta_{j}=\lambda_{j}$ into
the coefficients of $\lambda^{0}$ in (2.8), we obtain
$$\alpha_{j}=\lambda_{j}^{2}.$$

Thus we obtain the Lax pair of the CHESCS (2.5)
\begin{subequations}
\label{eqns:laxpair}
 \begin{align}
&\varphi_{xx}=(\frac{1}{4}+\lambda q)\varphi,\\
& \varphi_{t}=\frac{u_{x}}{2}\varphi+(\frac{1}{2\lambda
}-u)\varphi_{x}+2 \sum\limits_{j=1}^{N}\frac{\lambda\lambda_{j}
\varphi_{j}}{\lambda-\lambda_{j}}(\varphi_{jx}\varphi-\varphi_{j}\varphi_{x}).
 \end{align}
\end{subequations}
which means that the CHESCS (2.5) is Lax integrable.

\section{The infinite conservation laws of the CHESCS}
\setcounter{equation}{0}With the help of the Lax representation of
the CHESCS, we could find the conservation laws for the CHESCS by a
well-known method. First we assume that $q$, $u$, $\varphi_j$ and
its derivatives tend to 0 when $|x|\rightarrow\infty$. Set
\begin{equation}
\Gamma=\frac{\varphi_x}{\varphi},
\end{equation}
then the identity
\begin{equation}
\frac{\partial}{\partial t}(\frac{\partial\ln\varphi}{\partial
x})=\frac{\partial}{\partial x}(\frac{\partial\ln\varphi}{\partial
t})\notag
\end{equation}
together with (2.10) implies that CHESCS has the following
conservation law:
\begin{equation}
\frac{\partial}{\partial t}(\Gamma)=\frac{\partial}{\partial
x}(\frac{\varphi_{t}}{\varphi})=\frac{\partial}{\partial
x}(\frac{1}{2}u_x+2\sum_{j=1}^N\frac{\lambda\lambda_j}{\lambda-\lambda_j}\varphi_j\varphi_{jx}+((\frac{1}{2\lambda}-u)-2\sum_{j=1}^N\frac{\lambda\lambda_j}{\lambda-\lambda_j}\varphi_j^2)\Gamma)
\end{equation}.
Using (2.10a) gives rise to
\begin{equation}
\Gamma_x=\frac{1}{4}+q\lambda-\Gamma^2.
\end{equation}
Let
\begin{equation}
\Gamma=\sum_{m=0}^\infty\mu_m\lambda^{\frac{1-m}{2}},
\end{equation}
then $\mu_m$ is the density of conservation laws.

Define
\begin{equation}
\frac{1}{2}u_x+2\sum_{j=1}^N\frac{\lambda\lambda_j}{\lambda-\lambda_j}\varphi_j\varphi_{jx}
+((\frac{1}{2\lambda}-u)-2\sum_{j=1}^N\frac{\lambda\lambda_j}{\lambda-\lambda_j}\varphi_j^2)\Gamma
=\sum_{m=0}^\infty F_m\lambda^{\frac{1-m}{2}}
\end{equation}

It is found that the density of the conservation laws $\mu_m$ and
the flux of the conservation laws $F_m$ satisfy the following
recursion relation:
\begin{eqnarray}
\mu_0&=&\sqrt{q},\notag\\
\mu_1&=&-\frac{1}{4}\frac{q_x}{q},\notag\\
\mu_2&=&\frac{1}{32}(\frac{4}{\sqrt{m}}+\frac{m_x^2}{m^{5/2}}-(\frac{4m_x}{m^{3/2}})_x),\notag\\
\mu_m&=&\frac{-\mu_{m-1,x}-\sum_{i=1}^{m-1}\mu_i\mu_{m-1-i}}{2\mu_0},\quad
m\geq3,
\end{eqnarray}
\begin{eqnarray}
F_0&=&(-u-2\sum_{j=1}^N\lambda_j\varphi_j^2)\sqrt{q},\notag\\
F_1&=&(u+2\sum_{j=1}^N\lambda_j\varphi_j^2)\frac{q_x}{4q}+\frac{1}{2}u_x+2\sum_{j=1}^N\lambda_j\varphi_j\varphi_{jx},\notag\\
F_{2m}&=&\sum_{i=0}^m(-u^{(i)}-2\sum_{j=1}^N\lambda_j^{i+1}\varphi_j^2)\mu_{2m-2i},\quad m\geq1,\notag\\
F_{2m+1}&=&\sum_{i=0}^m(u^{(i)}+2\sum_{j=1}^N\lambda_j^{i+1}\varphi_j^2)\mu_{2m-2i+1}+2\sum_{j=1}^N\lambda_j^{m+1}\varphi_j\varphi_{jx},\quad
m\geq1,
\end{eqnarray}
where $u^{(0)}=u$, $u^{(1)}=1$, $u^{(i)}=0,\quad i>1$.

After some calculations we can find the first few conserved
quantities given by $\mu_0$, $\mu_2$ and $\mu_4$ are as follows
\begin{subequations}
\begin{eqnarray}
H_{-1}&=&\int\sqrt{q}dx,\\
H_{-2}&=&-\frac{1}{16}\int(\frac{4}{\sqrt{q}}+\frac{q_x^2}{q^{5/2}})dx,\\
H_{-3}&=&-\int(\frac{1}{32q^{3/2}}+\frac{5q_x^2}{64q^{7/2}}+\frac{q_{xx}^2}{32q^{7/2}}-\frac{35q_x^4}{512q^{11/2}})dx.
\end{eqnarray}
\end{subequations}

The corresponding flux of the conservation laws are
\begin{subequations}
\begin{eqnarray}
G_{-1}&=&(-u-2\sum_{j=1}^N\lambda_j\varphi_j^2)\sqrt{q},\\
G_{-2}&=&(1+2\sum_{j=1}^N\lambda_j^2\varphi_j^2)\sqrt{q}+(u+2\sum_{j=1}^N\lambda_j\varphi_j^2)(\frac{1}{16}(\frac{4}{\sqrt{q}}+\frac{q_x^2}{q^{5/2}})-(\frac{q_x}{4q^{3/2}})_x),\\
G_{-3}&=&(-u-2\sum_{j=1}^N\lambda_j\varphi_j^2)(\frac{1}{32q^{3/2}}+\frac{5q_x^2}{64q^{7/2}}+\frac{q_{xx}^2}{32q^{7/2}}-\frac{35q_x^4}{512q^{11/2}})\notag\\
&+&\frac{1}{16}(1+2\sum_{j=1}^N\lambda_j^2\varphi_j^2)(\frac{4}{\sqrt{q}}+\frac{q_x^2}{q^{5/2}})+2\sum_{j=1}^N\lambda_j^3\varphi_j^2\sqrt{q}.
\end{eqnarray}
\end{subequations}

As the space part of the Lax Pair of the CHESCS is the same as that
of CH equation, the densities of the conservation laws of the CHESCS
are the same as those of the Camassa-Holm equation $^{12}$. As the
time part of the Lax pair is different, the fluxs of the
conservation laws for CH equation and CHESCS are different.

\section{One peakon solution of the CHESCS}
\setcounter{equation}{0} The CH equation (\ref{eq:CH}) has peakon
solutions $^{2}$
\begin{equation}\label{eq:peak}
u=ce^{-|x-ct+\alpha|},
\end{equation}
where $\alpha$ is an arbitrary constant. The corresponding
eigenfunction of (2.1) is
\begin{equation}
\varphi=\beta e^{-\frac{1}{2}|x-ct+\alpha|},
\end{equation}
where $\beta$ is an arbitrary constant.

Since the CHESCS (2.5) can be considered as the CH equation
(\ref{eq:CH}) with non-homogeneous terms, we may use the method of
variation of constants to find the peakon solution of CHESCS from
the peakon solution (4.1) and (4.2). Taking $\alpha$ and $\beta$ in
(\ref{eq:peak}) and (4.2) to be time-dependent $\alpha(t)$ and
$\beta(t)$ and requiring that
\begin{subequations}
\begin{eqnarray}\label{eq:CHESpeak}
u&=&ce^{-|x-ct+\alpha(t)|},\\
\varphi&=&\beta(t)e^{-\frac{1}{2}|x-ct+\alpha(t)|}
\end{eqnarray}
\end{subequations}
satisfy the CHESCS (2.5) for $N=1$. We find that
$c=\frac{1}{\lambda}$, $\alpha(t)$ can be an arbitrary function of
$t$ and $\beta(t)=\sqrt{\alpha'(t)c}$. So we have the one peakon
solution for (2.4) with $N=1$, $\lambda_1=\lambda=\frac{1}{c}$
\begin{subequations}
\begin{eqnarray}
u&=&ce^{-|x-ct+\alpha(t)|}\\
\varphi&=&\sqrt{\alpha'(t)c}e^{-\frac{1}{2}|x-ct+\alpha(t)|}
\end{eqnarray}
\end{subequations}

The one peakon of the CHESCS also has a cusp at its peak, located at
$x=ct-\alpha(t)$. We note that for the one peakon solution of the CH
equation, the solution travels with speed $c$ and has a cusp at its
peak of height $c$, for the CHESCS, the cusp is still at its peak of
height $c$, but the speed $c-\frac{\alpha(t)}{t}$ of the wave is no
longer a constant.

\section{A reciprocal transformation for the CHESCS}
\setcounter{equation}{0}Let $r=\sqrt{q}$, by the reciprocal
transformation $^{4,28,29}$
 $$dy=r dx-urds,~ds=dt,$$
and denoting $f=r^{-\frac{1}{2}}\phi$, the Lax pair (2.1) of CH
equation is transformed to the following system
\begin{subequations}
\label{eqns:newlaxpair}
  \begin{align}
&\phi_{yy}=(\lambda+Q+\frac{1}{4\omega})\phi,\\
&\phi_{s}=\frac{1}{2\lambda}(r\phi_{y}-\frac{1}{2}r_{y}\phi),
\end{align}
\end{subequations}
where
\begin{equation}
\label{eqns:vu}
Q=-\frac{1}{4}(\frac{r_{y}}{r})^{2}+\frac{r_{yy}}{2r}+\frac{1}{4r^{2}}-\frac{1}{4\omega}.
\end{equation}
The compatibility condition of (5.1a) and (5.1b) gives an
alternative of the associated CH (ACH) equation
\begin{subequations}
\label{eqns:system}
  \begin{align}
&Q_{s}=r_{y},\\
&-\frac{1}{4\omega}r_{y}+\frac{1}{4}r_{yyy}-\frac{1}{2}Q_{y}r-Q
r_{y}=0.
\end{align}
\end{subequations}
We now consider the reciprocal transformation for the CHESCS (2.5).
(2.5a) gives
\begin{equation}
\label{eqns:conservelaw}
r_{t}=-(ru)_{x}-2\sum\limits_{j=1}^{N}\lambda_{j}(r\varphi_{j}^{2})_{x}.
\end{equation}
(\ref{eqns:conservelaw}) shows that the 1-form
\begin{equation}
\label{eqns:1form} \omega=r
dx-(ru+2\sum\limits_{j=1}^{N}\lambda_{j}r\varphi_{j}^{2})dt
\end{equation}
is closed, so we can define a reciprocal transformation
$(x,t)\rightarrow (y,s)$ by the relation
\begin{equation}
\label{eqns:trans} dy=r
dx-(ru+2\sum\limits_{j=1}^{N}\lambda_{j}r\varphi_{j}^{2})ds,~ds=dt,
\end{equation}
and we have
\begin{equation}
\label{eqns:trans1} \frac{\partial}{\partial
x}=r\frac{\partial}{\partial y},~\frac{\partial}{\partial
t}=\frac{\partial}{\partial
s}-(ru+2\sum\limits_{j=1}^{N}\lambda_{j}r\varphi_{j}^{2})\frac{\partial}{\partial
y}.
\end{equation}
Denoting
$\varphi=r^{-\frac{1}{2}}\psi,~\varphi_{j}=r^{-\frac{1}{2}}\psi_{j}$
and using (\ref{eqns:vu}), the Lax pair (2.9) of CHESCS (2.5) is
correspondingly rewritten as
\begin{subequations}
\label{eqns:newlaxpair1}
  \begin{align}
&\psi_{yy}=(\lambda+Q+\frac{1}{4\omega})\psi,\\
&\psi_{s}=\frac{1}{2\lambda}(r\psi_{y}-\frac{1}{2}r_{y}\psi)+2\sum\limits_{j=1}^{N}
\frac{\lambda_{j}^{2}
\psi_{j}}{\lambda-\lambda_{j}}(\psi_{jy}\psi-\psi_{j}\psi_{y}).
\end{align}
\end{subequations}
The compatibility condition of (5.8a) and (5.8b) leads to an
associated CHESCS (ACHESCS)
\begin{subequations}
\label{eqns:newform}
  \begin{align}
&
Q_{s}=r_{y}-8\sum\limits_{j=1}^{N}\lambda_{j}^{2}\psi_{j}\psi_{jy},\\
&-\frac{1}{4\omega}r_{y}+\frac{1}{4}r_{yyy}-\frac{1}{2}Q_{y}r-Q
r_{y}=0,\\
&\psi_{jyy}=(\lambda_{j}+Q+\frac{1}{4\omega})\psi_{j},~j=1,2,\cdots,N.
\end{align}
\end{subequations}
The Eqs.(\ref{eqns:newform}) can be regarded as the
Eqs.(\ref{eqns:system}) with self-consistent sources. In order to
obtain the solutions of the CHESCS (2.5), we have to get the
relation of the variables $(y,s)$ and the variables $(x,t)$. From
the reciprocal transformation, we have
\begin{equation}
\label{eqns:pde} \frac{\partial x}{\partial y}=\frac{1}{r},~
\frac{\partial x}{\partial
s}=u+2\sum\limits_{j=1}^{N}\lambda_{j}\varphi_{j}^{2}.
\end{equation}
By making use of the compatibility of the above two equations, we
have
\begin{equation}
\label{eqns:pde} x(y,s)=\int\frac{1}{r}dy.
\end{equation}
The solutions of the CHESCS (2.5) with respect to the variables
(y,s) are given by
\begin{subequations}
\label{eqns:solution}
\begin{align}
 & q=r^{2}(y,s),~\varphi_{j}(y,s)=\frac{\psi_{j}}{\sqrt{r}},\\
 & u(y,s)=r^{2}
-r_{ys}+\frac{r_{y}r_{s}}{r}-2rr_{y}\sum\limits_{j=1}^{N}\lambda_{j}(\varphi_{j}^{2})_{y}
-2r^{2}\sum\limits_{j=1}^{N}\lambda_{j}(\varphi_{j}^{2})_{yy}-\omega,\\
&x(y,s)=\int\frac{1}{r}dy.
\end{align}
\end{subequations}
We now prove (5.12b). From $q=u-u_{xx}+\omega$
 and the reciprocal transformation (\ref{eqns:trans1}), we have
\begin{equation}
\label{eqns:u} u=q+rr_{y}u_{y}+r^{2}u_{yy}-\omega.
\end{equation}
By using the reciprocal transformation (\ref{eqns:trans1}),
(\ref{eqns:conservelaw}) gives rise to
\begin{equation}
\label{eqns:uy}
u_{y}=-\frac{r_{s}}{r^{2}}-2\sum\limits_{j=1}^{N}\lambda_{j}(\varphi_{j}^{2})_{y}.
\end{equation}
Substituting (\ref{eqns:uy}) and (\ref{eqns:vu}) into (\ref{eqns:u})
leads to (5.12b).

\section{The solutions for the CHESCS}
\setcounter{equation}{0}Notice that $Q=0$, $r=\sqrt{\omega}$ is the
solution of (5.2) and (5.3). Let the functions
$\phi_0(y,s,\lambda)$, $\Psi_1(y,s,\lambda_1)$,
$\cdots,\Psi_n(y,s,\lambda_n)$ be different solutions of (5.1) with
$Q=0$, $r=\sqrt{\omega}$ and the corresponding $\lambda$ and
$\lambda=\lambda_1,\cdots,\lambda_n$, respectively. We construct two
Wronskian determinants from these functions
\begin{subequations}
\begin{eqnarray}
W_1=W(\Psi_1,\cdots,\Psi_1^{(m_1)},\Psi_2,\cdots,\Psi_2^{(m_2)},\cdots,\Psi_n,\cdots,\Psi_n^{(m_n)}),\\
W_2=W(\Psi_1,\cdots,\Psi_1^{(m_1)},\Psi_2,\cdots,\Psi_2^{(m_2)},\cdots,\Psi_n,\cdots,\Psi_n^{(m_n)},\phi_0),
\end{eqnarray}
\end{subequations}
where $m_i\geq0$ are given numbers and
$\Psi_j^{(i)}=\frac{\partial^i\Psi_j(y,s,\lambda)}{\partial\lambda^i}|_{\lambda=\lambda_j}$.

Based on the generalized Darboux transformation for KdV hierarchy
$^{30}$ and using (5.3a), the following generalized Darboux
transformation of (5.1) is valid $^{4,30,31}$
\begin{subequations}
\begin{eqnarray}
Q(y,s)=-2\frac{\partial^2}{\partial y^2}\log W_1,\\
r(y,s)=\sqrt{\omega}-2\frac{\partial^2}{\partial y\partial s}\log
W_1,\\
\phi(y,s,\lambda)=\frac{W_2}{W_1},
\end{eqnarray}
\end{subequations}
namely $Q(y,s)$, $r(y,s)$ and $\phi(y,s,\lambda)$ satisfy (5.1),
(5.2) and (5.3).

\subsection{The multisoliton solutions}
Take $\Psi_{i}$ and $\Phi_{i}$ be the solutions of Eq.(5.1) with
$Q=0,~$ $r=\sqrt{\omega}$ and
 $~\lambda_{i}=k_{i}^{2}-\frac{1}{4\omega}<0$, or $4\omega
 k_i^2-1<0$, $(0<k_{1}<k_{2}<\cdots<k_{n})$ as follows
\begin{subequations}
\begin{eqnarray}
&\Psi_{i}=cosh\xi_{i}, & ~ i ~~is~ an~ odd ~number,\\
&\Psi_{i}=sinh\xi_{i}, & ~i~~ is~ an ~even ~number.
\end{eqnarray}
\end{subequations}
\begin{equation}
\label{eqns:phi}\Phi_{i}=e^{\xi_i}
 \end{equation}
where henceforth
\begin{equation}
\label{eqns:xi} \xi_{i}=k_{i}[y+\frac{2\omega^{3/2}s}{4\omega
k_{i}^{2}-1}+\alpha_{i}].
\end{equation}
By using Darboux transformation (6.2) with $m_1=\cdots=m_n=0$, the
n-soliton solution $Q(y,s)$ and $r(y,s)$ of (5.3) and the
corresponding eigenfunction $\phi_i(y,s,\lambda_i)$ of (5.1) with
$\lambda_i=k_i^2-\frac{1}{4\omega}$ is given by
\begin{subequations}
\label{eqns:dt}
  \begin{align}
&Q(y,s)=-2[logW(\Psi_{1},\Psi_{2},\cdots,\Psi_{n})]_{yy},\\
&r(y,s)=\sqrt{\omega}-2[logW(\Psi_{1},\Psi_{2},\cdots,\Psi_{n})]_{ys},\\
&\phi_{i}(y,s,\lambda_{i})=\frac{W(\Psi_{1},\Psi_{2},\cdots,\Psi_{n},\Phi_{i})}{W(\Psi_{1},\Psi_{2},\cdots,\Psi_{n})}.
 \end{align}
\end{subequations}
When $n=1$ and $4k_1^2\omega-1<0$, (6.6) gives rise to one soliton
solution for (5.3) and the corresponding eigenfunction of (5.1) with
$\lambda_1=k_1^2-\frac{1}{4\omega}$ $^{8,9}$
\begin{subequations}
\label{eqns:ones}
  \begin{align}
&Q(y,s)=-2k_{1}^{2}sech^{2}\xi_{1},~r(y,s)=\sqrt{\omega}-\frac{4k_{1}^{2}\omega^{\frac{3}{2}}sech^{2}\xi_{1}}
{4k_{1}^{2}\omega-1},\\
&\phi_{1}=k_{1}sech\xi_{1}.
\end{align}
\end{subequations}

Since Eq.(\ref{eqns:newform}) can be considered to be
Eq.(\ref{eqns:system}) with non-homogeneous terms and $\phi_{1}$
satisfies (5.1a) with $\lambda=\lambda_{1}$, we may apply the method
of variation of constant to find the solutions of the
 CHESCS (\ref{eqns:newform}) by using the solution (\ref{eqns:ones})
 of ACH equation (\ref{eqns:system}) and corresponding eigenfunction. Taking $\alpha_{1}$ in (6.5) to be time-dependent functions $\alpha_{1}(s)$
and requiring that
\begin{subequations}
\label{eqns:ones1}
  \begin{align}
&\bar{Q}(y,s)=-2k_{1}^{2}sech^{2}\bar{\xi}_{1},~\bar{r}(y,s)=\sqrt{\omega}-\frac{4k_{1}^{2}
\omega^{\frac{3}{2}}sech^{2}\bar{\xi}_{1}}
{4k_{1}^{2}\omega-1},\\
&\bar{\psi}_{1}=\beta_1(s)k_{1}sech\bar{\xi}_{1}
\end{align}
\end{subequations}
satisfy the system (\ref{eqns:newform}) for $N=1$, henceforth, we
denote
\begin{equation}
\label{eqns:kx} \bar{\xi}_{i}=k_{i}[y+\frac{2\omega^{3/2}s}{4\omega
k_{i}^{2}-1}+\alpha_{i}(s)].
\end{equation}
We find that $\alpha_{1}(s)$ can be an arbitrary  function of $s$
and
\begin{equation}
\label{eqns:alpha}
\beta_{1}(s)=\frac{2\omega}{1-4k_{1}^{2}\omega}\sqrt{2\alpha_{1}'(s)}.
\end{equation}
So the one-soliton solution of the CHESCS (2.5) with $N=1$ and
$\lambda_1=k_1^2-\frac{1}{4\omega}<0$ is obtained with respect to
the variables $(y,s)$ from
 (\ref{eqns:solution})
\begin{subequations}
  \begin{align}
&q(y,s)=\omega(1-\frac{4k_{1}^{2}\omega
sech^{2}\bar{\xi}_{1}}{4k_{1}^{2}\omega-1})^{2},\\
&u(y,s)=\frac{8k_{1}^{2}\omega^{2}sech^{2}\bar{\xi}_{1}}{(1-4k_{1}^{2}\omega)
(1-4k_{1}^{2}\omega+4k_{1}^{2}\omega sech^{2}\bar{\xi}_{1})},\\
&\varphi_{1}(y,s)=\frac{2\sqrt{2\alpha_1'(s)}k_{1}\omega
sech\bar{\xi}_{1}}{\sqrt{\sqrt{\omega}(1-4k_{1}^{2}\omega)(1-4k_{1}^{2}\omega+4k_{1}^{2}\omega
sech^{2}\bar{\xi}_{1})}},\\
&x(y,s)=\frac{y}{\sqrt{\omega}}-2\ln\frac{1-2k_1\sqrt{\omega}\tanh\bar{\xi_1}}{1+2k_1\sqrt{\omega}\tanh\bar{\xi_1}}.
\end{align}
\end{subequations}
The requirement $4k_1^2\omega-1<0$ guarantees the nonsingularity of
solution (6.11).

In Fig 1, we plot the single soliton solution of $u$ and
$\varphi_1$.

\vskip 105pt
\begin{center}
\begin{picture}(35,25)
\put(-150,0){\resizebox{!}{3.3cm}{\includegraphics{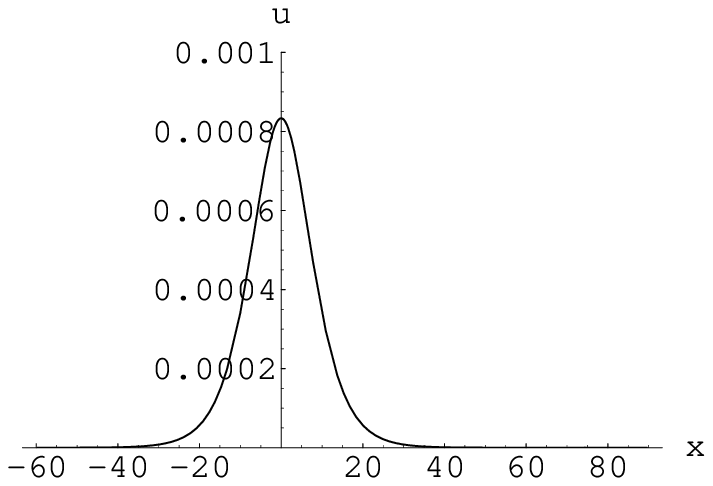}}}
\put(60,0){\resizebox{!}{3.3cm}{\includegraphics{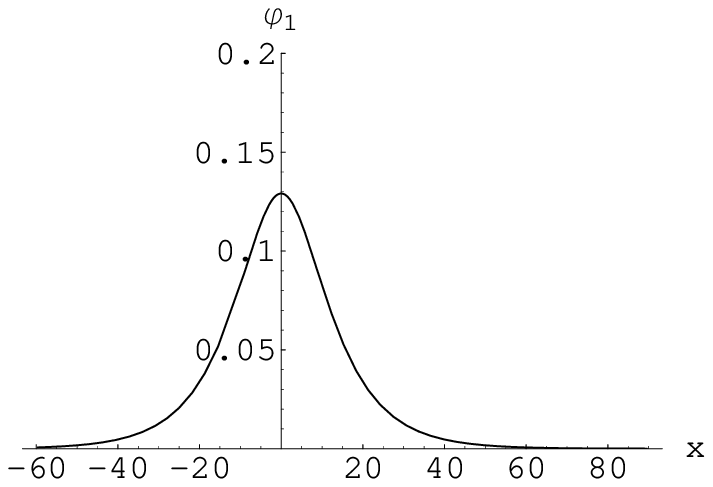}}}
\end{picture}
\end{center}
\begin{center}
\begin{minipage}{ 14cm}{\footnotesize
\vskip 10pt {\bf Figure 1.} Single soliton solutions for $u$ and the
eigenfunction $\varphi_{1}$ when $w=0.01,~k_{1}=1,
~\alpha_{1}(s)=4s,~ s=2.$ }
\end{minipage}
\end{center}

When $n=2$, $\lambda_1=k_1^2-\frac{1}{4\omega}<0$,
$\lambda_2=k_2^2-\frac{1}{4\omega}<0$, we have
\begin{subequations}
\begin{align}
&\Psi_1=cosh\xi_1,\quad\Psi_2=sinh\xi_2,\\
&\Phi_1=e^{\xi_1},\quad\Phi_2=e^{\xi_2},\\
&W_1(\Psi_1,\Psi_2)=k_2cosh\xi_2cosh\xi_1-k_1sinh\xi_2sinh\xi_1,\\
&W_2(\Psi_1,\Psi_2,\Phi_1)=k_2(k_2^2-k_1^2)sinh\xi_1,\\
&W_2(\Psi_1,\Psi_2,\Phi_2)=k_1(k_1^2-k_2^2)cosh\xi_2,\\
\end{align}
\end{subequations}
Then (6.6) with $n=2$ gives rise to two soliton solution for (5.3)
and the corresponding eigenfunction of (5.1). In the same way as we
did on the one-soliton solution, we can apply the method of
variation of constants to get the two soliton solution of the
ACHESCS (2.9) which together with (5.12) yields to the two soliton
solution for CHESCS (2.5) with
$N=2,\lambda_1=k_1^2-\frac{1}{4\omega},\lambda_2=k_2^2-\frac{1}{4\omega}$
\begin{subequations}
  \begin{align}
  &r(y,s)=\sqrt{\omega}-2[logW_{1}(\Psi_{1},\Psi_{2})]_{ys}|_{\xi_{i}=\bar{\xi}_{i}},\\
  &\psi_{i}=\frac{2\omega \sqrt{(-1)^{i+1}2\alpha_{i}'(s)} W_{2}(\Psi_{1},\Psi_{2},\Phi_{i})}
  {(1-4k_{i}^{2}\omega)\sqrt{\prod\limits_{j\neq
  i}{(k_j^2-k_i^2)}}W_{1}(\Psi_{1},\Psi_{2})}|_{\xi_{i}=\bar{\xi}_{i}},i=1,2.
\end{align}
\end{subequations}
In Fig 2 we plot the interactions of two soliton solution for $u$
and $\varphi_1$, $\varphi_{2}$, which is shown that $u$ is elastic
collision.

\vskip 105pt
\begin{center}
\begin{picture}(35,25)
\put(-220,0){\resizebox{!}{3.3cm}{\includegraphics{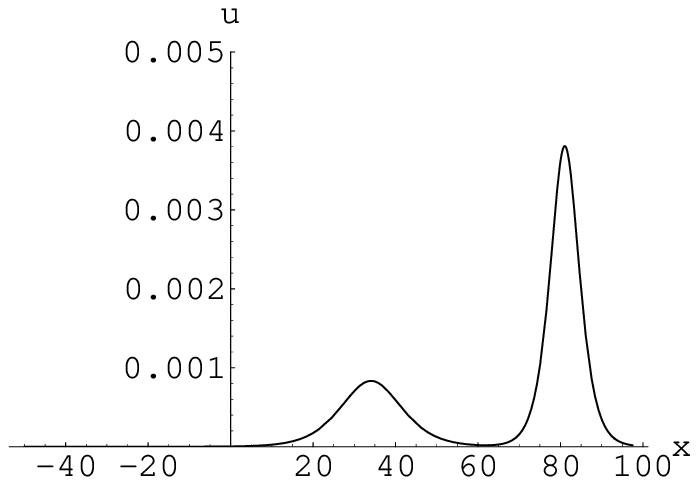}}}
\put(-60,0){\resizebox{!}{3.3cm}{\includegraphics{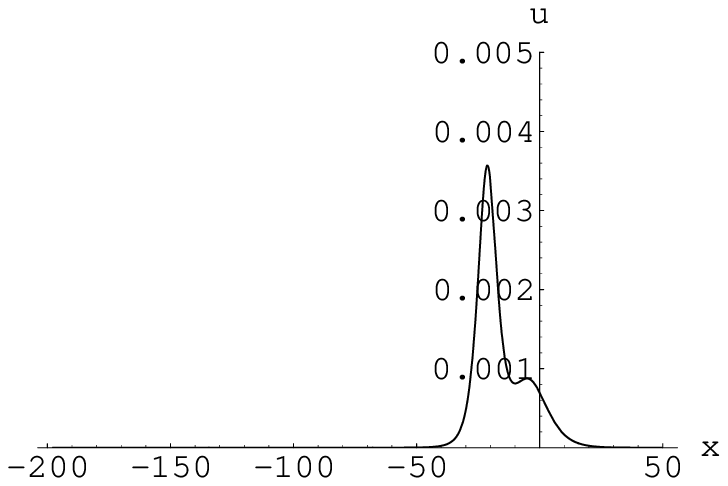}}}
\put(80,0){\resizebox{!}{3.3cm}{\includegraphics{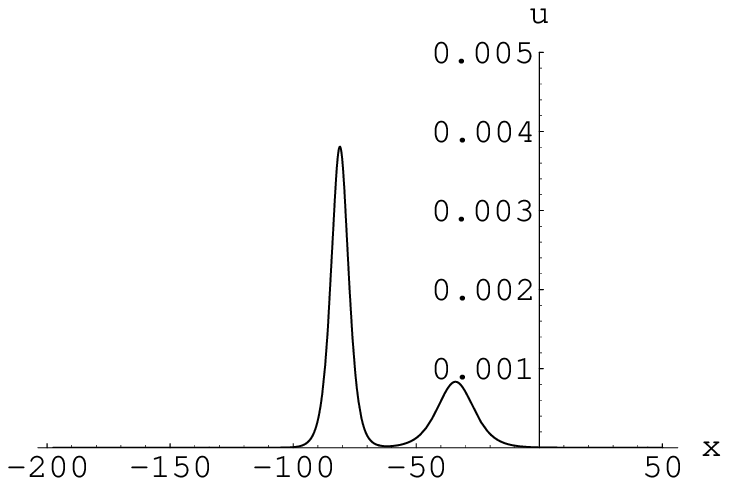}}}
\end{picture}
\end{center}

\begin{center}
\begin{minipage}{ 14cm}{\footnotesize
~~~~~~~~~~~~\quad\quad\quad\quad\quad(a)s=-2~~~~~\quad\quad\quad\quad\quad\quad\quad\quad\quad\quad\quad\quad\quad~~~~~~~~
~~~~~~~~~~~~~~~~~~~~~~~~~~~~~~~~~~~~~~~~(b)s=1~~~~~~~\quad\quad\quad\quad\quad\quad\quad\quad\quad\quad\quad\quad~~~~~~~~~~~
~~~~~~~~~~~~~~~~~~~~~~~~~~~~~~~~~~~~~~~~(c)s=2}
\end{minipage}
\end{center}

\begin{center}
\begin{picture}(35,25)
\put(-220,-50){\resizebox{!}{3.3cm}{\includegraphics{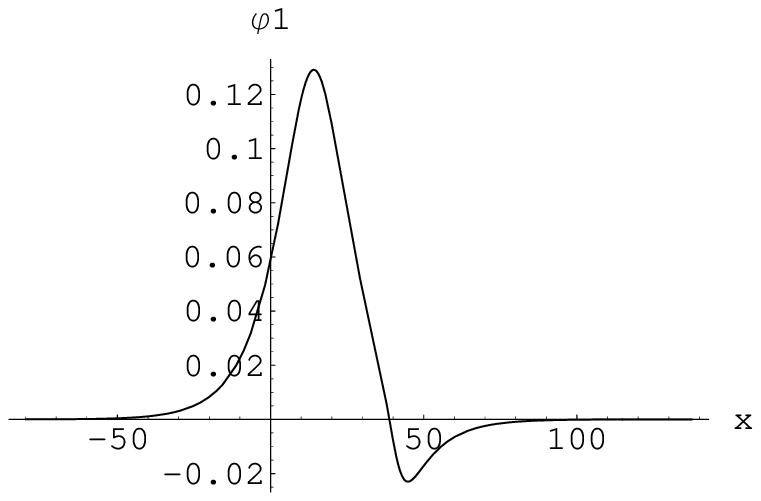}}}
\put(-60,-50){\resizebox{!}{3.3cm}{\includegraphics{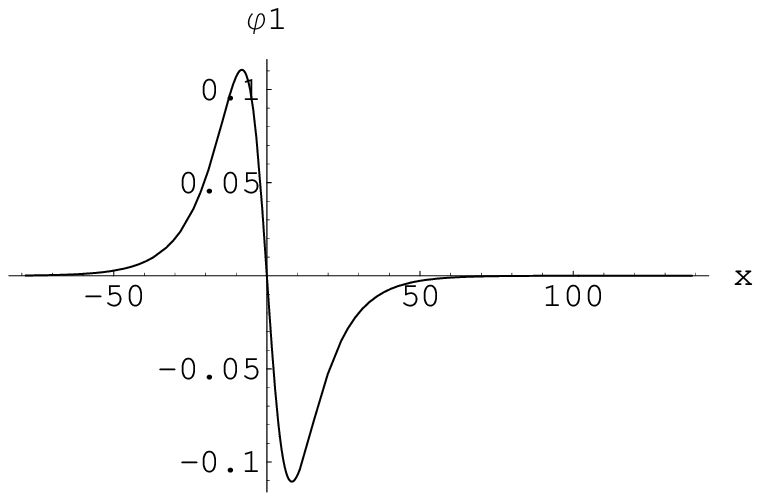}}}
\put(80,-50){\resizebox{!}{3.3cm}{\includegraphics{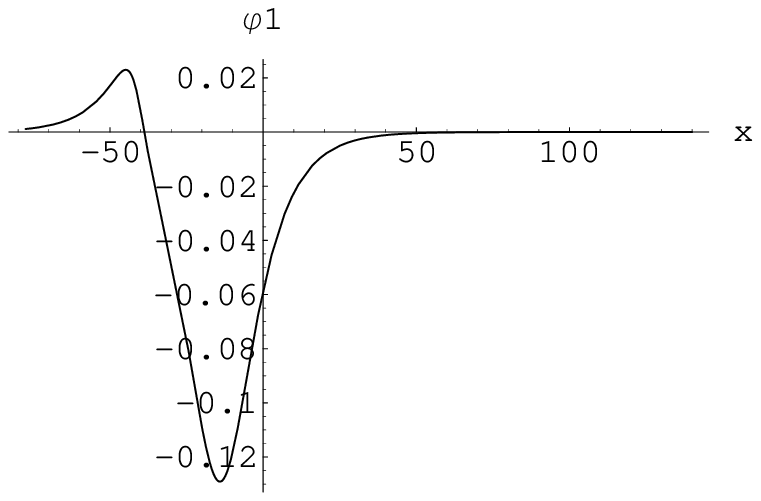}}}
\end{picture}
\end{center}

\vskip 30pt
\begin{center}
\begin{minipage}{ 20cm}{\footnotesize
~~~~~~~~~~~~~~~~~~~~~~\quad\quad\quad\quad\quad\quad(d)s=-1~~~~~~~~\quad\quad\quad\quad\quad\quad\quad\quad\quad\quad\quad\quad\quad~~~~~
~~~~~~~~~~~~~~~~~~~~~~~~~~~~~~~~~~~~~~~~(e)s=0~~~~~~~~\quad\quad\quad\quad\quad\quad\quad\quad\quad\quad\quad\quad~~~~~~~~~~~
~~~~~~~~~~~~~~~~~~~~~~~~~~~~~~~~~~~~~~~~(f)s=1}
\end{minipage}
\end{center}

\begin{center}
\begin{picture}(35,25)
\put(-220,-50){\resizebox{!}{3.3cm}{\includegraphics{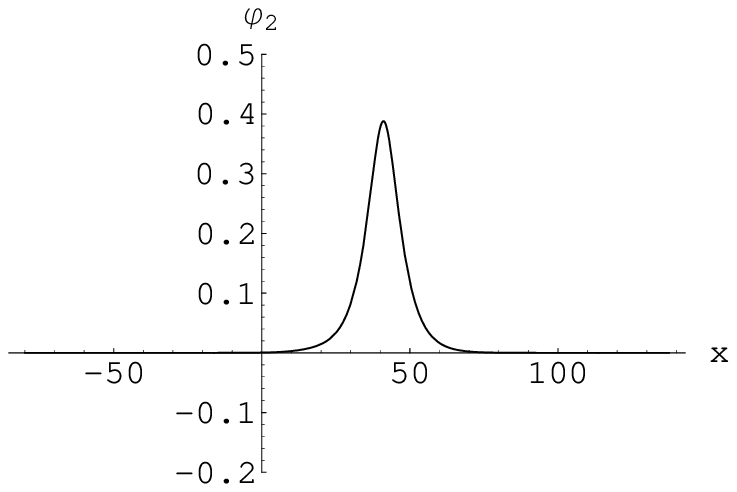}}}
\put(-60,-50){\resizebox{!}{3.3cm}{\includegraphics{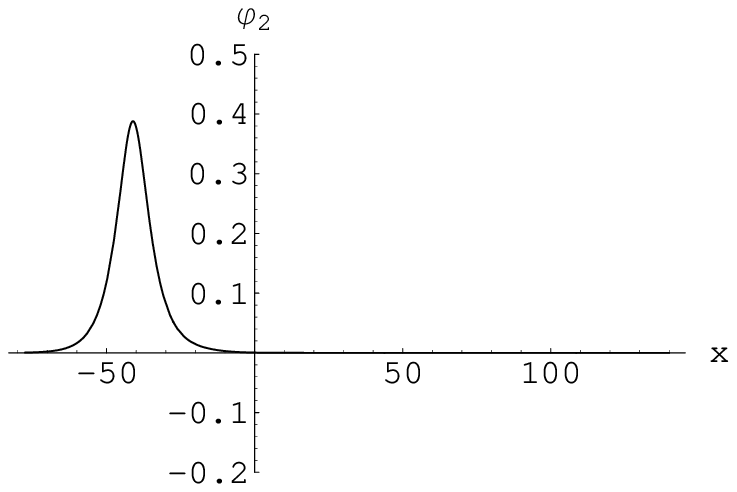}}}
\put(80,-50){\resizebox{!}{3.3cm}{\includegraphics{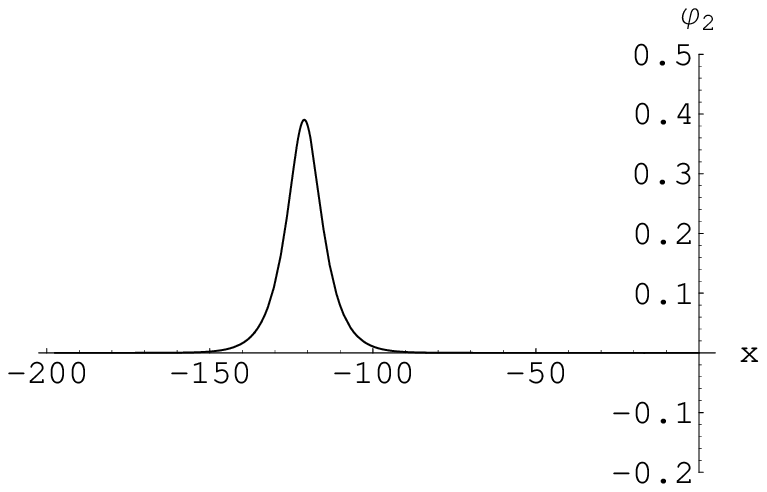}}}
\end{picture}
\end{center}

\vskip 30pt
\begin{center}
\begin{minipage}{ 30cm}{\footnotesize
~~~~~~~~~~~~~~~~~~~~~~\quad\quad\quad\quad\quad\quad(g)s=-1~~~~~~~~\quad\quad\quad\quad\quad\quad\quad\quad\quad\quad\quad\quad\quad~~~~~
~~~~~~~~~~~~~~~~~~~~~~~~~~~~~~~~~~~~~~~~(h)s=1~~~~~~~~\quad\quad\quad\quad\quad\quad\quad\quad\quad\quad\quad\quad~~~~~~~~~~~
~~~~~~~~~~~~~~~~~~~~~~~~~~~~~~~~~~~~~~~~(i)s=3\\
\vskip 15pt {\bf Figure 2.} two soliton solutions for $u$ and the
eigenfunction $\varphi_{1}$, $\varphi_{2}$ when
$w=0.01,~k_{1}=2,~k_{2}=1, ~\alpha_{1}(s)=2s,~\alpha_{2}(s)=4s.$ }
\end{minipage}
\end{center}

Notice that the soliton solutions of CHESCS contains arbitrary $s$
functions $\alpha_j(s)$. This implies that the insertion of sources
into the CH equation may cause the variation of the speed of
soliton.

In the same way as in the reference $^{27}$, we may apply the method
of variation of constant to find the N-soliton solution of (2.5)
with $\lambda_i=k_i^2-\frac{1}{4\omega}>0,\quad i=1,\cdots,N,$ from
(\ref{eqns:solution}), where
\begin{subequations}
\label{eqns:ns}
  \begin{align}
  &r(y,s)=\sqrt{\omega}-2[logW_{1}(\Psi_{1},\Psi_{2},\cdots,\Psi_{N})]_{ys}|_{\xi_{i}=\bar{\xi}_{i}},\\
  &\psi_{i}=\frac{2\omega \sqrt{(-1)^{i+1}2\alpha_{i}'(s)} W_{2}(\Psi_{1},\Psi_{2},\cdots,\Psi_{N},\Phi_{i})}
  {(1-4k_{i}^{2}\omega)\sqrt{\prod\limits_{j\neq
  i}{(k_j^2-k_i^2)}}W_{1}(\Psi_{1},\Psi_{2},\cdots,\Psi_{N})}|_{\xi_{i}=\bar{\xi}_{i}}
\end{align}
\end{subequations}

\subsection{The multicuspon solutions}
Take $\Psi_{i}$ and $\Phi_{i}$ be the solutions of
Eq.(\ref{eqns:newlaxpair}) when $Q=0,~$ $r=\sqrt{\omega}$ and
 $~\lambda_{i}=k_{i}^{2}-\frac{1}{4\omega}>0$
$(0<k_{1}<k_{2}<\cdots<k_{n})$, as follows
\begin{subequations}
\begin{eqnarray}
&\Psi_{i}=sinh\xi_{i}, & ~ i ~~is~ an~ odd ~number,\\
&\Psi_{i}=cosh\xi_{i}, & ~i~~ is~ an ~even ~number.
\end{eqnarray}
\end{subequations}
\begin{equation}
\Phi_{i}=e^{\xi_i},
 \end{equation}
where $\xi_{i}$ is given by (6.5).

The n-cuspon solution $Q(y,s)$ and $r(y,s)$ of (5.3) and the
corresponding eigenfunction $\phi_i(y,s,\lambda_i)$ of (5.1) with
$\lambda_i=k_i^2-\frac{1}{4\omega}$ is given by (6.6).

When $n=1$ and $4k_1^2\omega-1>0$, (6.6) gives rise to one cuspon
solution for (5.3) and the corresponding eigenfunction of (5.1) with
$\lambda_1=k_1^2-\frac{1}{4\omega}$ $^{8,9}$
\begin{subequations}
  \begin{align}
&Q(y,s)=2k_{1}^{2}csch^{2}\xi_{1},~r(y,s)=\sqrt{\omega}+\frac{4k_{1}^{2}\omega^{\frac{3}{2}}csch^{2}\xi_{1}}
{4k_{1}^{2}\omega-1},\\
&\phi_{1}=-k_{1}csch\xi_{1}.
\end{align}
\end{subequations}

Similarly, we may apply the method of variation of constant to find
the solutions of the ACHESCS (5.9) by using the solution (6.17) of
(5.3) and corresponding eigenfunction. Taking $\alpha_1$ in (6.5) to
be time-dependent functions $\alpha_1(s)$ and requiring that
\begin{subequations}
\label{eqns:ones1}
  \begin{align}
&\bar{Q}(y,s)=2k_{1}^{2}csch^{2}\bar{\xi}_{1},~\bar{r}(y,s)=\sqrt{\omega}+\frac{4k_{1}^{2}
\omega^{\frac{3}{2}}csch^{2}\bar{\xi}_{1}}
{4k_{1}^{2}\omega-1},\\
&\bar{\psi}_{1}=\beta_1(s)k_{1}csch\bar{\xi}_{1}
\end{align}
\end{subequations}
satisfy the system (\ref{eqns:newform}) for $N=1$, we find that
$\alpha_1(s)$ can be an arbitrary function of $s$ and
\begin{equation}
\label{eqns:alpha}
\beta_{1}(s)=\frac{2\omega}{1-4k_{1}^{2}\omega}\sqrt{-2\alpha_{1}'(s)}.
\end{equation}
So the one-cuspon solution of the CHESCS (2.5) with $N=1$ and
$\lambda_1=k_1^2-\frac{1}{4\omega}>0$ is obtained with respect to
the variables $(y,s)$ from (5.12)
\begin{subequations}
\label{eqns:ones1}
\begin{align}
&q(y,s)=\omega(1+\frac{4k_{1}^{2}\omega
csch^{2}\bar{\xi}_{1}}{4k_{1}^{2}\omega-1})^{2},\\
&u(y,s)=\frac{8k_{1}^{2}\omega^{2}csch^{2}\bar{\xi}_{1}}{(1-4k_{1}^{2}\omega)
(-1+4k_{1}^{2}\omega+4k_{1}^{2}\omega csch^{2}\bar{\xi}_{1})},\\
&\varphi_{1}(y,s)=\frac{2\sqrt{2\alpha_1'(s)}k_{1}\omega
csch\bar{\xi}_{1}}{\sqrt{\sqrt{\omega}(1-4k_{1}^{2}\omega)(-1+4k_{1}^{2}\omega+4k_{1}^{2}\omega
csch^{2}\bar{\xi}_{1})}},\\
&x(y,s)=\frac{y}{\sqrt{\omega}}+2\ln\frac{1-2k_1\sqrt{\omega}\coth\bar{\xi_1}}{1+2k_1\sqrt{\omega}\coth\bar{\xi_1}}
\end{align}
\end{subequations}

In Fig 3, we plot the one-cuspon solution of $u$, $\varphi_1$.

\vskip 105pt
\begin{center}
\begin{picture}(35,25)
\put(-150,0){\resizebox{!}{3.3cm}{\includegraphics{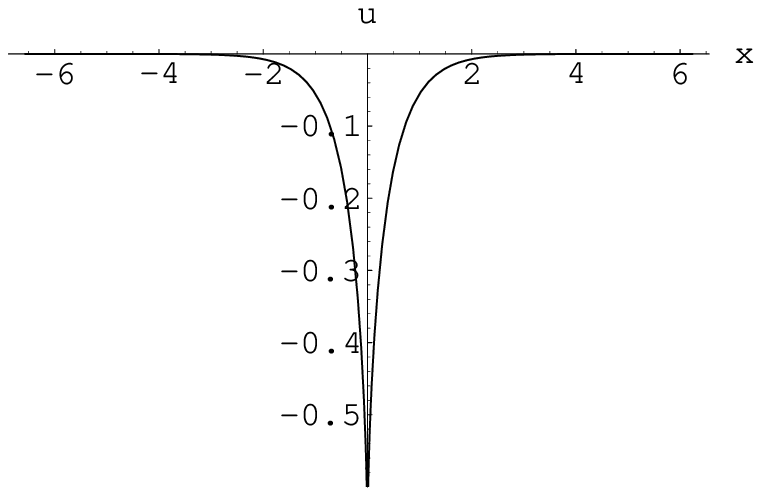}}}
\put(60,0){\resizebox{!}{3.3cm}{\includegraphics{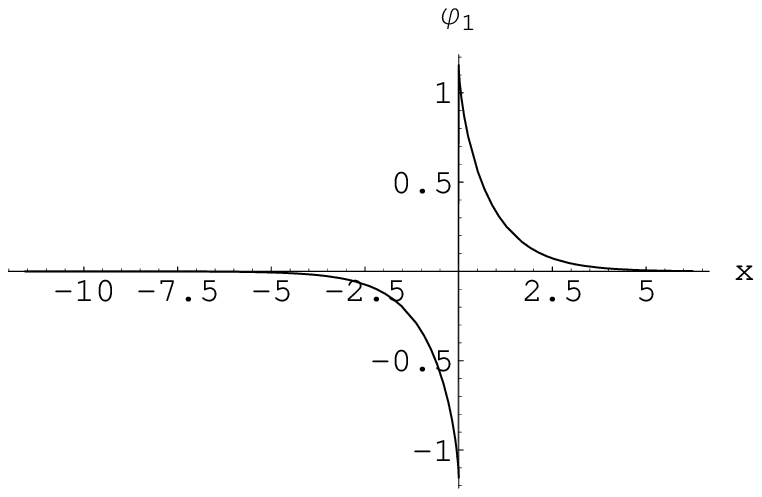}}}
\end{picture}
\end{center}
\begin{center}
\begin{minipage}{ 14cm}{\footnotesize
\vskip 10pt {\bf Figure 3.} Single cuspon solution for $u$ and the
eigenfunction $\varphi_{1}$ when $w=1,~k_{1}=1, ~\alpha_{1}(s)=-2s,~
s=2.$ }
\end{minipage}
\end{center}

Similarly, we can apply the method of variation of constant to find
the N-cuspon solution of (2.5) with
$\lambda_i=k_i^2-\frac{1}{4\omega}<0,\quad i=1,\cdots,N$ from
(\ref{eqns:solution}), where
\begin{subequations}
\label{eqns:ns}
  \begin{align}
  &r(y,s)=\sqrt{\omega}-2[logW_{1}(\Psi_{1},\Psi_{2},\cdots,\Psi_{N})]_{ys}|_{\xi_{i}=\bar{\xi}_{i}},\\
  &\psi_{i}=\frac{2\omega \sqrt{(-1)^{i+1}2\alpha_{i}'(s)} W_{2}(\Psi_{1},\Psi_{2},\cdots,\Psi_{N},\Phi_{i})}
  {(1-4k_{i}^{2}\omega)\sqrt{\prod\limits_{j\neq
  i}{(k_j^2-k_i^2)}}W_{1}(\Psi_{1},\Psi_{2},\cdots,\Psi_{N})}|_{\xi_{i}=\bar{\xi}_{i}}
\end{align}
\end{subequations}

Further more in the same way, we can fnd mixed
$k_1$-soliton-$k_2$-cuspon solution for (2.5) with $N=k_1+k_2$,
$\lambda_i=k_i^2-\frac{1}{4\omega}>0,\quad i=1,\cdots,k_1$ and
$\lambda_i=k_i^2-\frac{1}{4\omega}<0,\quad i=K_1+1,\cdots,k_1+k_2,$
by using (6.6) and (5.12).

\subsection{The multipositon solutions}
Let $\lambda=-k^2-\frac{1}{4\omega}$,
$\lambda_i=-k_i^2-\frac{1}{4\omega},\quad i=1,\cdots,N$, and take
\begin{subequations}
\begin{eqnarray}
&\Psi_{i}=sin\xi_{i}, & ~ i ~~is~ an~ odd ~number,\\
&\Psi_{i}=cos\xi_{i}, & ~i~~ is~ an ~even ~number.
\end{eqnarray}
\end{subequations}
\begin{subequations}
\begin{eqnarray}
&\Phi_{i}=cos\xi_{i}, & ~ i ~~is~ an~ odd ~number,\\
&\Phi_{i}=sin\xi_{i}, & ~i~~ is~ an ~even ~number.
\end{eqnarray}
\end{subequations}
where
$$\xi=k(y-\frac{2\omega^{3/2}s}{4k^2\omega+1})+\sum_{i=1}^N\prod_{j=1}^N(k-k_j)^2\frac{\alpha_i}{k-k_i},$$
$$\xi_i=\xi|_{k=k_i}.$$
For $N=1$, we have
\begin{subequations}
\begin{align}
&\Psi_1=\sin\xi_1,\quad\Psi_1^{(1)}=\gamma_1\cos\xi_1,\\
&\xi_1=k_1(y-\frac{\sqrt{\omega}s}{2(k_1^{2}+\frac{1}{4\omega})}).\\
&\gamma_1=\frac{\partial\xi}{\partial
k}|_{k=k_1}=\alpha_1+y+\frac{16k_1^2\omega^{5/2}s}{(1+4k_1^2\omega)^2}-\frac{2\omega^{3/2}s}{1+4k_1^2\omega},
\end{align}
\end{subequations}
and
\begin{subequations}
\begin{align}
&W_1(\Psi_1,\Psi_1^{(1)})=-k_1\gamma_1+\frac{1}{2}\sin{2\xi_1},\\
&W_2(\Psi_1,\Psi_1^{(1)},\Phi_1)=-2k_1^2\sin{\xi_1}.
\end{align}
\end{subequations}
Then the one-positon solution of (5.3) and the corresponding
eigenfunction for (5.1) is given by (6.2) with $N=1,m_1=1,$
\begin{subequations}
\begin{align}
&Q(y,s)=-2[logW_1]_{yy},\\
&r(y,s)=\sqrt{\omega}-2[logW_1]_{ys}.\\
&\psi_1(y,s,\lambda_1)=\beta_1\frac{W_2}{W_1},
\end{align}
\end{subequations}
where $\alpha_1$ and $\beta_1$ are arbitrary constants.

\vskip 105pt
\begin{center}
\begin{picture}(35,25)
\put(-150,0){\resizebox{!}{3.3cm}{\includegraphics{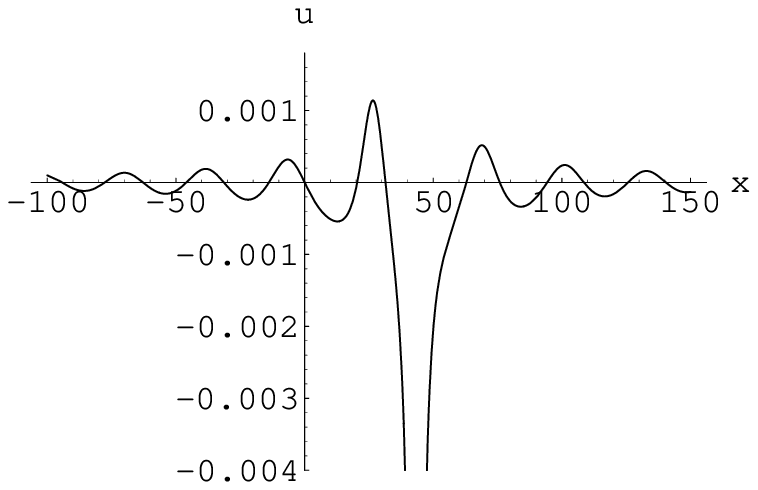}}}
\put(60,0){\resizebox{!}{3.3cm}{\includegraphics{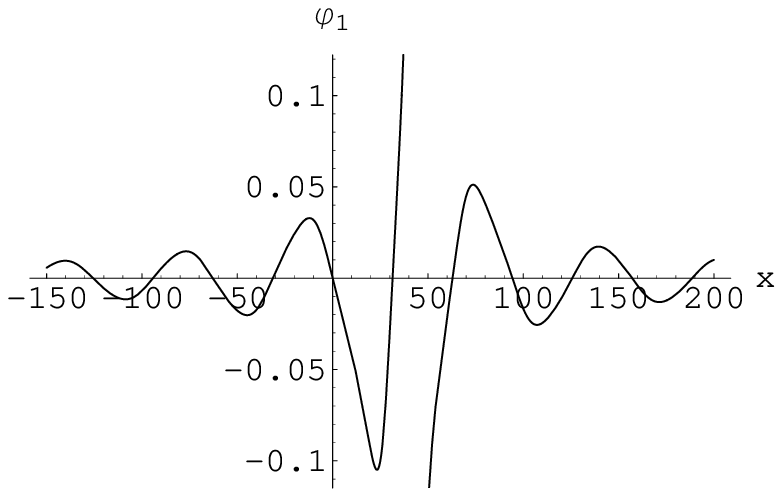}}}
\end{picture}
\end{center}
\begin{center}
\begin{minipage}{ 14cm}{\footnotesize
\vskip 10pt {\bf Figure 4.} one-positon solutions for $u$ and the
eigenfunction $\varphi_{1}$ when $w=0.01,~k_{1}=1,
~\alpha_{1}(s)=-2s,~ s=2.$ }
\end{minipage}
\end{center}

By using the method of variation of constants, which means we change
$\alpha_1$ and $\beta_1$ into $\alpha_1(s)$ and $\beta_1(s)$, we
obtain the one-positon solution for the CHESCS (2.5) with
$N=1,\lambda_1=-k_1^2-\frac{1}{4\omega}$ from (5.12), where
\begin{subequations}
\begin{align}
&\bar{r}(y,s)=\sqrt{\omega}-2[logW_1]_{ys}|_{\gamma_1=\bar{\gamma_1}},\\
&\bar{\psi_1}(y,s)=\frac{2\omega\sqrt{-\alpha_1'(s)}}{k_1(1+4k_1^2\omega)}\frac{W_2}{W_1}|_{\gamma_1=\bar{\gamma_1}},\\
&\bar{\gamma_1}=\alpha_1(s)+y+\frac{16k_1^2\omega^{5/2}s}{(1+4k_1^2\omega)^2}-\frac{2\omega^{3/2}s}{1+4k_1^2\omega}.
\end{align}
\end{subequations}
where $\alpha_1(s)$ is an arbitrary function of $s$.

In Fig 4, we plot the one-positon solution of $u$ and $\varphi_1$.

The positon solution of CHESCS is long-range analogue of soliton and
is slowly decreasing, oscillating solution $^{31}$. In the same way
we can find N-positon solution for (2.5). For a detailed discussion
on positon solution we refer to the reference $^{31}$.

For $N$, we have
\begin{subequations}
\begin{eqnarray}
&\Psi_{i}^{(1)}=\gamma_i\cos\xi_{i}, & ~ i ~~is~ an~ odd ~number,\\
&\Psi_{i}^{(1)}=-\gamma_i\sin\xi_{i}, & ~i~~ is~ an ~even ~number.
\end{eqnarray}
\end{subequations}
where
$$\gamma_i=\frac{\partial\xi}{\partial k}|_{k=k_i}=\prod_{j\neq i}(k_i-k_j)^2\alpha_i+y+\frac{16k_i^2\omega^{5/2}s}{(1+4k_i^2\omega)^2}-\frac{2\omega^{3/2}s}{1+4k_i^2\omega}.$$

We find that
\begin{subequations}
\begin{align}
&W_1=W(\Psi_1,\Psi_1^{(1)},\cdots,\Psi_N,\Psi_N^{(1)}),\\
&\phi_i=W(\Psi_1,\Psi_1^{(1)},\cdots,\Psi_N,\Psi_N^{(1)},\Phi_i),
\end{align}
\end{subequations}
and the N-positon solution of (5.3) and the corresponding
eigenfunction for (5.1) is given by
\begin{subequations}
\begin{align}
&Q(y,s)=-2[logW_1]_{yy},\\
&r(y,s)=\sqrt{\omega}-2[logW_1]_{ys}.\\
&\psi_j(y,s,\lambda_i)=\beta_i\frac{W_2}{W_1}, i=1,\cdots,N,
\end{align}
\end{subequations}
where $\alpha_i$ and $\beta_i$ are arbitrary constants.

By using the method of variation of constants, we obtain the
N-positon solution for the CHESCS (2.5) from (5.12), where
\begin{subequations}
\begin{align}
&\bar{r}(y,s)=\sqrt{\omega}-2[logW_1]_{ys}|_{\gamma_i=\bar{\gamma_i}},\\
&\bar{\psi_i}(y,s)=\frac{2\omega}{k_i(1+4k_i^2\omega)}\frac{1}{\prod_{j\neq i}{(k_j+k_i)}}\sqrt{(-1)\alpha'_i(s)}\frac{W_2}{W_1}|_{\gamma_i=\bar{\gamma_i}},\\
&\bar{\gamma_i}=\prod_{j\neq
i}(k_i-k_j)^2\alpha_i(s)+y+\frac{16k_i^2\omega^{5/2}s}{(1+4k_i^2\omega)^2}-\frac{2\omega^{3/2}s}{1+4k_i^2\omega}.
\end{align}
\end{subequations}
where $\alpha_i(s)$ are arbitrary functions of $s$.

\subsection{The multinegaton solutions}
Let $\lambda=k^2-\frac{1}{4\omega}>0$,
$\lambda_i=k_i^2-\frac{1}{4\omega}>0,\quad i=1,\cdots,N$, and take
\begin{subequations}
\begin{eqnarray}
&\Psi_{i}=sinh\xi_{i}, & ~ i ~~is~ an~ odd ~number,\\
&\Psi_{i}=cosh\xi_{i}, & ~i~~ is~ an ~even ~number.
\end{eqnarray}
\end{subequations}
\begin{equation}
\Phi_{i}=e^{\xi_i},
 \end{equation}
where
$$\xi=k(y+\frac{2\omega^{3/2}s}{4k^2\omega-1})+\sum_{i=1}^N\prod_{j=1}^N(k-k_j)^2\frac{\alpha_i}{k-k_i},$$
$$\xi_i=\xi|_{k=k_i},$$
then we have
\begin{subequations}
\begin{align}
&\Psi_1=\sinh\xi_1,\quad\Psi_1^{(1)}=\gamma_1\cosh\xi_1,\\
&\xi_1=k_1(y+\frac{\sqrt{\omega}s}{2(k_1^{2}-\frac{1}{4\omega})}).\\
&\gamma_1=\alpha_1+y+\frac{-16k_1^2\omega^{5/2}s}{(4k_1^2\omega-1)^2}+\frac{2\omega^{3/2}s}{4k_1^2\omega-1},
\end{align}
\end{subequations}
and
\begin{subequations}
\begin{align}
&W_1(\Psi_1,\Psi_1^{(1)})=-k_1\gamma_1+\frac{1}{2}\sinh{2\xi_1},\\
&W_2(\Psi_1,\Psi_1^{(1)},\Phi_1)=2k_1^2\sinh{\xi_1},
\end{align}
\end{subequations}
Then the one-negaton solution of (5.3) and the corresponding
eigenfunction for (5.1) is given by
\begin{subequations}
\begin{align}
&Q(y,s)=-2[logW_1]_{yy},\\
&r(y,s)=\sqrt{\omega}-2[logW_1]_{ys}.\\
&\phi_1(y,s,\lambda_1)=\beta_1\frac{W_2}{W_1},
\end{align}
\end{subequations}
where $\alpha$ and $\beta$ are arbitrary constants.

By using the method of variation of constants, we obtain the
one-negaton solution for the CHESCS (2.5) with
$N=1,\lambda_1=k_1^2-\frac{1}{4\omega}$ from (5.12), where
\begin{subequations}
\begin{align}
&\bar{r}(y,s)=\sqrt{\omega}-2[logW_1]_{ys}|_{\gamma_1=\bar{\gamma_1}},\\
&\bar{\psi_1}(y,s)=\frac{2\omega\sqrt{\alpha_1'(s)}}{k_1(4k_1^2\omega-1)}\frac{W_2}{W_1}|_{\gamma_1=\bar{\gamma_1}},\\
&\bar{\gamma_1}=\alpha_1(s)+y+\frac{-16k_1^2\omega^{5/2}s}{(4k_1^2\omega-1)^2}+\frac{2\omega^{3/2}s}{4k_1^2\omega-1}.
\end{align}
\end{subequations}
where $\alpha(s)$ is an arbitrary function of $s$.

In Fig 5, we plot the one-negaton solution of $u$ and $\varphi_1$.

\vskip 105pt
\begin{center}
\begin{picture}(35,25)
\put(-150,0){\resizebox{!}{3.3cm}{\includegraphics{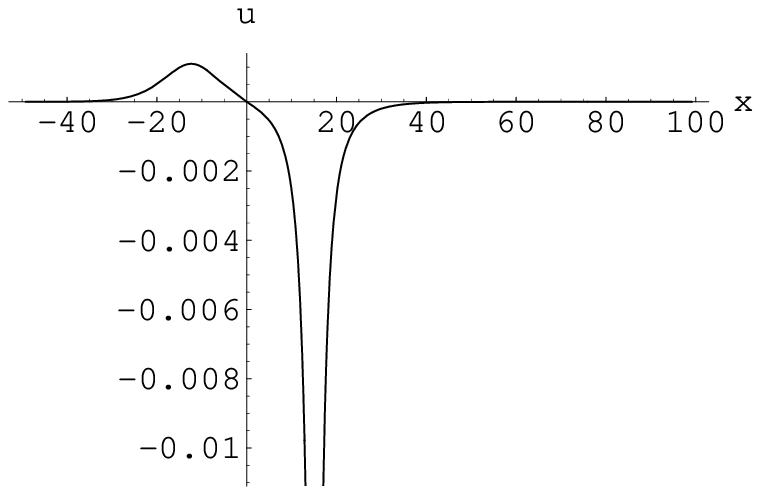}}}
\put(60,0){\resizebox{!}{3.3cm}{\includegraphics{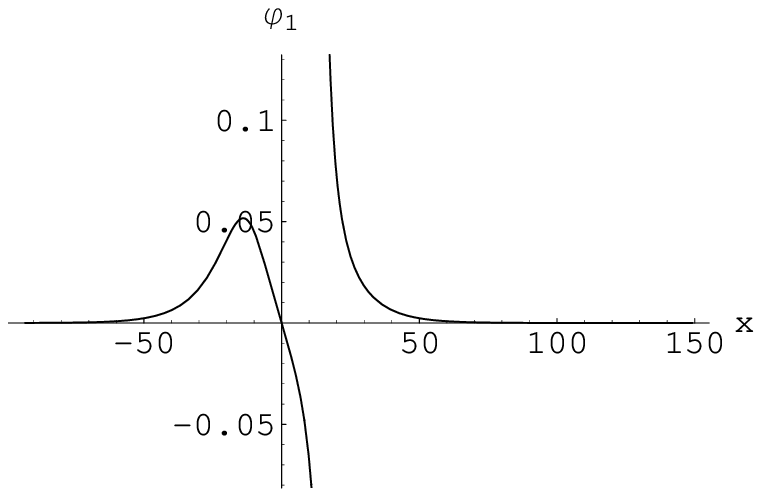}}}
\end{picture}
\end{center}
\begin{center}
\begin{minipage}{ 14cm}{\footnotesize
\vskip 10pt {\bf Figure 5.} one-negaton solution for $u$ and the
eigenfunction $\varphi_{1}$ when $w=0.01,~k_{1}=1,
~\alpha_{1}(s)=2s,~ s=2.$ }
\end{minipage}
\end{center}

Similarly, we can find N-negaton solution for CHESCS (2.5) with
$\lambda_i=k_i^2-\frac{1}{4\omega},\quad i=1,\cdots,N$. For $N$, we
have
\begin{subequations}
\begin{eqnarray}
&\Psi_{i}^{(1)}=\gamma_i\cosh\xi_{i}, & ~ i ~~is~ an~ odd ~number,\\
&\Psi_{i}^{(1)}=\gamma_i\sinh\xi_{i}, & ~i~~ is~ an ~even ~number.
\end{eqnarray}
\end{subequations}
where
$$\gamma_i=\frac{\partial\xi}{\partial k}|_{k=k_i}=\prod_{j\neq i}(k_i-k_j)^2\alpha_i+y+\frac{-16k_i^2\omega^{5/2}s}{(4k_i^2\omega-1)^2}+\frac{2\omega^{3/2}s}{4k_i^2\omega-1}.$$

We find that
\begin{subequations}
\begin{align}
&W_1=W(\Psi_1,\Psi_1^{(1)},\cdots,\Psi_N,\Psi_N^{(1)}),\\
&\phi_i=W(\Psi_1,\Psi_1^{(1)},\cdots,\Psi_N,\Psi_N^{(1)},\Phi_i),
\end{align}
\end{subequations}
and the N-negaton solution of (5.3) and the corresponding
eigenfunction for (5.1) is given by
\begin{subequations}
\begin{align}
&Q(y,s)=-2[logW_1]_{yy},\\
&r(y,s)=\sqrt{\omega}-2[logW_1]_{ys}.\\
&\psi_j(y,s,\lambda_i)=\beta_i\frac{W_2}{W_1}, i=1,\cdots,N,
\end{align}
\end{subequations}
where $\alpha_i$ and $\beta_i$ are arbitrary constants.

By using the method of variation of constants, we obtain the
N-negaton solution for the CHESCS (2.5) from (5.12), where
\begin{subequations}
\begin{align}
&\bar{r}(y,s)=\sqrt{\omega}-2[logW_1]_{ys}|_{\gamma_i=\bar{\gamma_i}},\\
&\bar{\psi_i}(y,s)=\frac{2\omega}{k_i(4k_i^2\omega-1)}\frac{1}{\prod_{j\neq i}{(k_j+k_i)}}\sqrt{\alpha'_i(s)}\frac{W_2}{W_1}|_{\gamma_i=\bar{\gamma_i}},\\
&\bar{\gamma_i}=\prod_{j\neq
i}(k_i-k_j)^2\alpha_i(s)+y+\frac{-16k_i^2\omega^{5/2}s}{(4k_i^2\omega-1)^2}+\frac{2\omega^{3/2}s}{4k_i^2\omega-1}.
\end{align}
\end{subequations}
where $\alpha_i(s)$ are arbitrary functions of $s$.

\section{Conclusion}
The CHESCS and its Lax representation are derived. Conservation laws
are constructed. It is reasonable to speculate on the potrential
application of CHESCS, that is, CHESCS may describe the interaction
of different solitary waves in shallow water. Since SESCS can be
regarded as soliton equations with non-homogeneous terms, we look
for explicit solutions by using the method of variation of
constants. By considering a reciprocal transformation, which relates
CH equation to an alternative of ACH equation, we propose a similar
reciprocal transformation, which relates the CHESCS to ACHESCS. By
using the Darboux transformation, one can find the n-soliton and
n-cuspon solution as well as n-positon and n-negaton solution of
alternative ACH equation. Then by means of the method of variation
of constants, we can obtain N-soliton, N-cuspon, N-positon and
N-negaton solutions of the ACHESCS. Finally, using the inverse
reciprocal transformation, we obtain N-soliton, N-cuspon, N-positon
and N-negaton solutions of the CHESCS.

\section*{Acknowledgements}
This work was supported by the National Basic Research Program of
China (973 program) \\(2007CB814800), China Postdoctoral Science
Foundation funded project (20080430420) and the National Science
Foundation of China (Grant no 10801083).

\section*{Reference}
$^{1}$ B. Fuchssteiner and A.S. Fokas, Physica D $\bf{4}$, 47
(1981).\\
$^{2}$ R. Camassa and D. Holm, Phys. Rev. Lett. $\bf{71}$, 1661
(1993).\\
$^{3}$ R. Camassa, D. Holm and J.Hyman, Adv. Appl. Mech.
$\bf{31}$, 1 (1994).\\
$^{4}$ A. Parker, Proc. R. Soc. Lond. A $\bf{460}$, 2929 (2004).\\
$^{5}$ R. S. Johnson, Proc. R. Soc. Lond. A $\bf{459}$, 1687 (2003).\\
$^{6}$ Z. J. Qiao, Commun. Math. Phys. $\bf{239}$, 309 (2003).\\
$^{7}$ R. Beals, D. H. Sattinger and J. Szmigielski, Adv. Math.
$\bf{154}$, 229 (2000).\\
$^{8}$ Y. S. Li and J. E. Zhang, Proc. R. Soc. Lond. A $\bf{460}$,
2617 (2004).\\
$^{9}$ Y. S. Li, J. Nonlinear Math. Phys. $\bf{12}$, 466 (2005).\\
$^{10}$ Z. J. Qiao and G. P. Zhang, EuroPhys. Lett. $\bf{73}$, 657
(2006).\\
$^{11}$ H. Holden, J. Hyp. Diff. Equ. $\bf{4}$, 39 (2007).\\
$^{12}$ J. Lenells, J. Phys. A: Math. Gen. $\bf{38}$, 869 (2005).\\
$^{13}$ V. K. Mel'nikov, Phys. Lett. A $\bf{133}$, 493 (1988).\\
$^{14}$ V. K. Mel'nikov, Commun. Math. Phys. $\bf{120}$, 451 (1989).\\
$^{15}$ V. K. Mel'nikov, Commun. Math, Phys. $\bf{126}$, 201 (1989).\\
$^{16}$ D. J. Kaup, Phys. Rev. Lett. $\bf{59}$, 2063 (1987).\\
$^{17}$ J. Leon and A. Latifi, J. Phys. A: Math. Gen. $\bf{23}$,
1385 (1990).\\
$^{18}$ C. Claude, A. Latifi and J. Leon, J. Math. Phys. $\bf{32}$,
3321 (1991).\\
$^{19}$ M. Nakazawa, E. Yomada and H. Kubota, Phys. Rev. Lett.
$\bf{66}$, 2625 (1991).\\
$^{20}$ E. V. Doktorov and V. S. Shchesnovich, Phys. Lett. A
$\bf{207}$, 153 (1995).\\
$^{21}$ Y. B. Zeng, Physica D $\bf{73}$, 171 (1994).\\
$^{22}$ Y. B. Zeng, W. X. Ma and Y. J. Shao, J. Math. Phys.
$\bf{42}$, 2113 (2001).\\
$^{23}$ Y. B. Zeng, Y. J. Shao, W. M. Xue, J. Phys. A: Math. Gen.
$\bf{36}$, 5035 (2003).\\
$^{24}$ T. Xiao and Y. B. Zeng, J. Phys. A: Math. Gen. $\bf{37}$,
7143 (2004).\\
$^{25}$ R. L. Lin, Y. B. Zeng and W. X. Ma, Physica A, $\bf{291}$,
287 (2001).\\
$^{26}$ H. X. Wu, Y. B. Zeng and T. Y. Fan, Inverse Problems
$\bf{24}$, 1 (2008).\\
$^{27}$ H. X. Wu, Y. B. Zeng, X. J. Liu and Y. H. Huang, Solving
soliton equations with self-consistent sources by constant variation
method, in submission.\\
$^{28}$ J. Schiff, Physica D $\bf{121}$, 24 (1998).\\
$^{29}$ A. Hone, J. Phys. A $\bf{32}$, L307 (1999).\\
$^{30}$ V. B. Matveev and M. A. Salle, Darboux transformations and
solitons, Springer-Verlag, 1991.\\
$^{31}$ R. Ivanov, Phys. Lett. A, $\bf{345}$, 112 (2005).
\end{document}